# ANI-1, A data set of 20 million calculated off-equilibrium conformations for organic molecules

Justin S. Smith[1], Olexandr Isayev[2,*], Adrian E. Roitberg[1,*]

[1]*Department of Chemistry, University of Florida, Gainesville, FL 32611, USA*

[2]*UNC Eshelman School of Pharmacy, University of North Carolina at Chapel Hill, Chapel Hill, NC 27599, USA*

* Corresponding authors; email: OI (olexandr@olexandrisayev.com) and AER (roitberg@ufl.edu)

## Abstract

One of the grand challenges in modern theoretical chemistry is designing and implementing approximations that expedite *ab initio* methods without loss of accuracy. Machine learning (ML) methods are emerging as a powerful approach to constructing various forms of transferable atomistic potentials. They have been successfully applied in a variety of applications in chemistry, biology, catalysis, and solid-state physics. However, these models are heavily dependent on the quality and quantity of data used in their fitting. Fitting highly flexible ML potentials, such as neural networks, comes at a cost: a vast amount of reference data is required to properly train these models. We address this need by providing access to a large computational DFT database, which consists of more than 20M off equilibrium conformations for 57,462 small organic molecules. We believe it will become a new standard benchmark for comparison of current and future methods in the ML potential community.

Design Type(s):          *in silico* molecular design, database creation, benchmarks
Measurement Type(s):     computational chemistry
Technology Type(s):      quantum chemistry, computational modeling
Factor Type(s):          level of theory
Sample Characteristic(s): Cartesian coordinates, total electronic energy

## Background and summary

Accurate descriptions of atomic and intermolecular interactions are a cornerstone of reliable computer simulations in biophysics, chemistry, and materials science. For the past 50 years we have seen tremendous progress in the development of theoretical methods and software tools aiming to describe more complex systems and allow for longer time scales. Kohn-Sham density-functional theory (KS-DFT or DFT for short) has become by far the most popular electronic structure method in computational physics and chemistry.[1] DFT has found applications in many systems in organic chemistry[2,3], biology[4], catalysis[3,5] and solid state chemistry[6,7]. It is also frequently combined with molecular dynamics (AIMD) and classical force fields (quantum mechanics–molecular mechanics (QM-MM)) to describe chemical reactions in extended systems.

Although DFT calculations have become affordable on modern supercomputers, we face a dilemma: standard computational algorithms representing the N electrons system require $O(N^2)$ storage and $O(N^3)$ arithmetic operations. This $O(N^3)$ complexity has become a critical bottleneck which limits capabilities to study larger realistic physical systems, as well as longer time scales relevant to actual experiment. Consequently, a lot of progress has been made in the development of atomistic potentials using machine learning (ML).[8,9] The low numerical complexity and high accuracy of machine learning algorithms makes them very attractive as a pragmatic substitute for *ab-initio* and DFT methods. Thanks to their remarkable ability to find complex relationships among data, in many cases these 'machine learned' models out-perform more physically sound approximations (like force fields) and methods while also reducing the computational time required for a given application.[9–15] These models are heavily dependent on the quality and quantity of data used in their fitting, also called training. Neural networks are highly efficient and effective at modeling reference training data, due to their flexible functional form. However, this flexibility comes at a cost: a vast amount of reference data is required to properly train these models.

The Chemical Space Project[16] computationally enumerated all possible organic molecules up to a certain size, resulting in the creation of the GDB databases. Their latest GDB-17 database[17] contains 166.4 billion molecules of up to 17 atoms of C, N, O, S, and halogens. All molecules follow the valency rules and are filtered for unstable substructures, non-synthesizable and strained topologies. GDB molecules are stored as SMILES [*www.opensmiles.org*] strings representing the composition and connectivity of a molecule.

The GDB databases were fundamental in creating the QM7 dataset[18], one of the first benchmark datasets for training atomistic ML potentials. The QM7 dataset consists of 7,165 energy minimized (equilibrium) molecules calculated with the PBE0 functional. All structures are a small subset of GDB-13 (older GDB database of nearly 1 billion organic molecules) composed of molecules with up to 7 heavy atoms C, N, O, and S. Later, QM7 was extended to include 13 additional properties, like frontier molecular orbital energies, dipole moments, polarizability, and excitation energies[19]. The first ML model trained on QM7 used kernel ridge regression with the Coulomb matrix representation, which predicted atomization energies with a mean absolute error (MAE) of 9.9 kcal × mol$^{-1}$. This error was quickly reduced to 3.3 kcal × mol$^{-1}$ [20] and eventually was under 1 kcal × mol$^{-1}$ [21].

QM9, is perhaps the most well-known benchmark dataset.[17,22] It consists of 133,885 equilibrium organic molecules containing up to nine heavy atoms (CONF) from the GDB-17 database. In addition to energy minima it reports corresponding harmonic frequencies, dipole moments, polarizabilities, along with energies, enthalpies, and free energies of atomization. All properties were calculated at the B3LYP/6-31G(2df,p) level of quantum chemistry. A subset of 6,095 constitutional isomers in QM9 corresponding to a brutto formula C7H10O2 was also calculated at the more accurate G4MP2 level of theory. Various molecular representations and ML methods were benchmarked against the QM9 dataset.[23,20,21,24] See also a recent survey of methods.[23] Later, a Message Passing Neural Network (MPNN)[10] achieved chemical accuracy in 11 out of 13 target properties in the QM9 dataset. Finally, the hierarchical interacting particle neural network (HIP-NN)[15] model of Lubbers et. al. achieved state-of-the-art accuracy of just 0.26 kcal × mol$^{-1}$ MAE on total energy prediction.

A common feature of all QMx datasets is that they only explore chemical degrees of freedom by providing information about energy minimized (equilibrium) molecular configurations. In these molecules, the forces of all atoms are equal to zero. Therefore, considerable efforts were undertaken to produce off-equilibrium datasets using *ab initio* molecular dynamics (AIMD) simulations. The C7O2H10-17 dataset includes energies from AIMD trajectories of 113 isomers of C7O2H10 (5k frames each). All simulations used the DFT/PBE level of theory and were carried out at 500 K. Very recently Schutt et al.[21] and Chmiela et al.[25] released MD17 dataset, a collection of eight AIMD/ PBE+vdW-TS simulations for small organic molecules. Each of these consist of an MD trajectory for a single molecule extending from ~100K to 900K frames. In contrast to the QMx datasets, these MD datasets explore conformational space while keeping composition fixed.

We recently introduced a neural network potential (NNP) called ANI-1, the first NNP for organic molecules shown to transfer to molecular systems well outside of its training set. As presented, the ANI-1 potential was trained on a data set, which spans **both conformational and configurational space**, built from small organic molecules of up to 8-heavy atoms. We show its applicability to much larger systems, up to 50 atoms, including well known drug molecules and a random selection of molecules from the GDB-11[26,27] database with 10-heavy atoms. ANI-1 shows exceptional predictive power on the 10-heavy atom test set, with RMSE versus DFT relative energies as low as 0.57 kcal/mol when only considering molecular conformations that are within 30 kcal/mol of the energy minimum for each molecule. More recently, Gastegger et. al.[28], showed similar results for large organic systems that were fragmented into smaller molecules and DFT data was generated on the fly for training. This was done in an active-learning fashion where the goal is to train the potential to a specific system during an MD simulation. Shortly after, Huang and Von Lilienfeld[29] used a fragmentation scheme for training an ML model to predict energies of large rigid drug molecules. Both studies back up the argument that information about the physics of large systems can be learned from data sets of small molecules.

In this data descriptor, we report a large dataset of non-equilibrium DFT total energy calculations for organic molecules. In total, we provide access to the total energies of ~20M molecular conformations for 57,462 molecules from the GDB database[26,27], which samples both chemical and conformational degrees of freedom at the same time. As the accuracy of modern ML methods for molecules in equilibrium on the QM9 benchmark achieved 1 kcal × mol$^{-1}$, ANI-1 provides 100x more data and a much more challenging task to learn. Therefore, we expect it will become a new standard benchmark of comparison for current and future methods in the machine learned potential community. More importantly, it is a sound foundation for the development of future general-purpose machine learned potentials, providing an exhaustive head start on data generation, which can be augmented with future data sets covering relevant regions of chemical space.

## Methods

**QM Calculations**. All electronic structure calculations are carried out with the ωB97x[30] density functional and the 6-31G(d) basis set[31] in the Gaussian 09[32] electronic structure package. ωB97x is a hybrid-meta GGA functional[30], which has been shown to be chemically accurate compared to high-level CCSD(T) calculations.[33–37]

**Molecular geometry generation.** The GDB-11 database[26,27] provides an exhaustive search of stable and chemically viable molecules, supplied in the SMILES [*www.opensmiles.org*] string format, containing C, N, O, and F atoms with up to 11 of these 'heavy' atoms. Hydrogen atoms are added through the RDKit cheminformatics software package [www.rdkit.org] to make molecular structures that are charge neutral and have a singlet electronic ground state. The ANI-1 data set presented here is built from an exhaustive sampling of a subset of the GDB-11 database containing molecules with between 1 and 8 heavy atoms and limiting the atomic species to C, N, and O. This leaves a subset of 57,947 starting molecules. All molecules are neutral and with a singlet electronic ground state. The conformation generation process is carried out in five steps starting with these 57,947 molecules. The steps are listed below and qualitatively depicted in Figure 1.

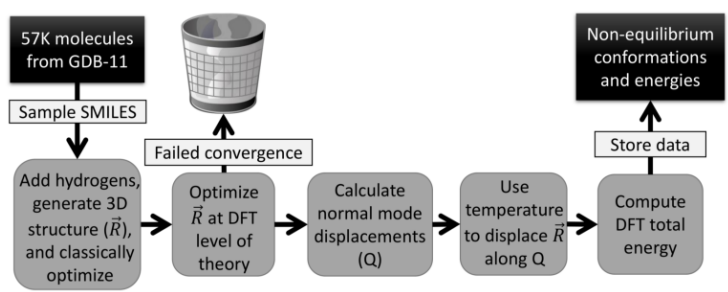

*Figure 1*: **Schematic representation of data generation.** Scheme for generating non-equilibrium conformations of 57,462 molecules from the GDB-11 database. The goal with this scheme is to generate a "window" of the potential surface around each optimized equilibrium structure.

Smiles strings from the GDB-11 subset described above are used to generate 3D conformations using RDKit. Also with RDKit, all structures are saturated with hydrogens such that each has charge 0 and multiplicity 1. The 3D structures are then pre-optimized to a stationary point using the MMFF94 force field[38] as implemented in RDKit.

At the chosen DFT or *ab-initio* level of theory, geometries are optimized until energy minima convergence. Optimization is carried out using Gaussian 09's default method and convergence criteria. Obtained geometries correspond to the first stationary point reached on the potential surface and correspond to some local minima or in a rare case to a saddle point. If convergence fails, the structure is not included in the data set. At this step, 485 (0.84% of total) molecules failed to converge during the structural optimization. The final data set is built from these 57,462 equilibrium geometries. Finally, for each of the 57,462 structurally optimized molecules, a normal mode calculation is performed in the Gaussian 09 package to obtained normal mode coordinates and their associated force constants. This is accomplished using the UltraFine DFT grid option with the ωB97x density functional.

**Normal Mode Sampling (NMS)**. To carry out normal mode sampling on an energy minimized molecule of $N_a$ atoms, first a set of $N_f$ normal mode coordinates, $Q = \{q_1, q_2, q_3, \ldots q_{N_f}\}$, is computed at the desired *ab-initio* level of theory, where $N_f = 3N_a - 5$ for linear molecules and $N_f = 3N_a - 6$ for all others. The corresponding force constants, $K = \{K_1, K_2, K_3, \cdots, K_{N_f}\}$, are obtained alongside $Q$. Then a set of $N_f$ uniformly distributed pseudo-random numbers, $c_i$, are generated such that $\sum_i^{N_f} c_i$ is in the range [0,1]. Next, a displacement, $R_i$, for each normal mode coordinate is computed by setting a harmonic potential equal to the $c_i$ scaled average energy of the system of particles at some temperature, T. Solving for the displacement gives,

$$R_i = \pm \sqrt{\frac{3c_i N_a k_b T}{K_i}} \qquad (5)$$

where $k_b$ is Boltzmann's constant. The sign of $R_i$ is determined randomly from a Bernoulli distribution where $p = 0.5$ to ensure that both sides of the harmonic potential are sampled equally. Each $R_i$ is used to scale the normalized normal mode coordinates by $q_i^R = R_i q_i$. Next, a new conformation of the molecule is generated by displacing the structurally optimized coordinates by $Q^R$, the superposition of all $q_i^R$. Finally, a single point energy at the desired level of theory is calculated using the newly displaced coordinates as input.

| Number of heavy atoms | Total Molecules | Max Temperature | S value | Energies < 275kcal/mol | Energies > 275kcal/mol | Total data points |
|---|---|---|---|---|---|---|
| 1 | 3 | 2,000.00 | 500 | 10,800 | 0 | 10,800 |
| 2 | 13 | 1,500.00 | 450 | 50,962 | 398 | 51,360 |
| 3 | 20 | 1,000.00 | 425 | 151,200 | 0 | 151,200 |
| 4 | 61 | 600 | 400 | 651,936 | 6,144 | 658,080 |
| 5 | 267 | 600 | 200 | 1,813,151 | 9,889 | 1,823,040 |
| 6 | 1,406 | 600 | 30 | 1,682,245 | 29,963 | 1,712,208 |
| 7 | 7,760 | 600 | 20 | 6,460,162 | 869,222 | 7,329,384 |
| 8 | 47,932 | 450 | 5 | 11,236,918 | 1,714,819 | 12,951,737 |
| Total | 57,462 | - | - | 22,057,374 | 2,630,435 | 24,687,809 |

**Table 1:** List of information and parameters used to generate the ANI-1 data set. The first column represents the number of heavy atoms per molecule in the test set. Total represents a combination of all sets. The molecules are obtained from the GDB-11 database. Column descriptions: "Total Molecules" is the number of molecules for each set, "Max Temperature" is the maximum temperature used to randomly perturb the molecule along the normal modes, "S value" is the number of data points generated per degree of freedom, and "Total data points" is the number of non-equilibrium structures generated for each set. The number of structures that fall below and above 275kcal/mol from the lowest energy structure in each molecule's set of conformers is also included.

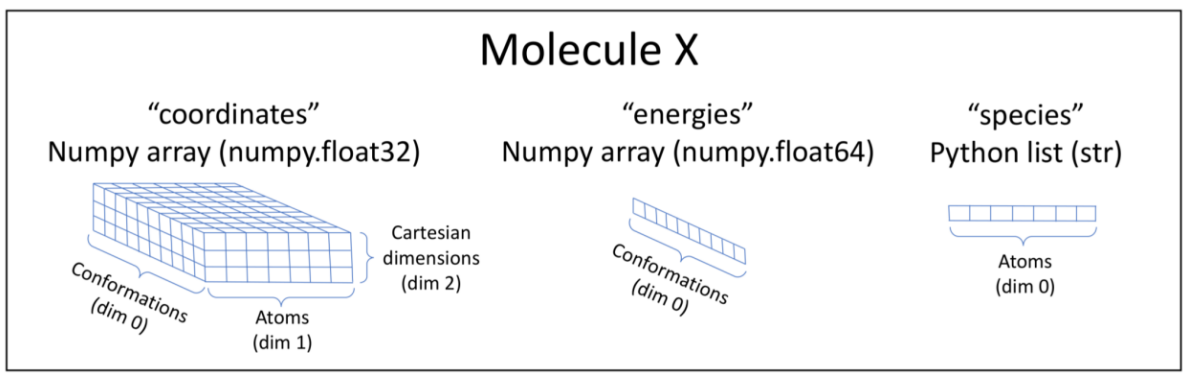

**Figure 2: Data Structure.** *Description of the containers stored in the dictionary returned by iterating through molecules stored in the HDF5 file format. The 'coordinates' key gives access to a 3D array containing each conformer of the molecule in cartesian coordinates, while the 'energies' key gives the 1D array of energies for the conformers. The first dimension of each 'coordinates' and 'energies' array maps correctly to the corresponding structure. The 'species' key contains the atomic symbol of the atoms and is ordered to correspond to the correct atoms in the second dimension of the array returned by the 'coordinates' key. Other keys in the returned dictionary are: "coordinatesHE", "energiesHE", and "smiles" for the high energy coordinates, high energy energies and SMILES string, respectively.*

N data points (new conformations) are generated, representing a window of the potential surface. N is calculated by $S \times K$ where *S* is an empirically chosen value (See Table 1) based on the number of heavy atoms in each molecule and *K* is the number of degrees of freedom of the molecule. The total energy, atomic symbols, and cartesian coordinates of the structure are stored as described in the Data Format section.

## Data Records

The data set is provided in an HDF5 based file in a Figshare data repository (Data Citation 1). A GitHub repository containing a README file with technical usage details and examples of how to access the data set is supplied online. [https://github.com/isayev/ANI1_dataset]

**File format**

Data is stored per molecule as described in figure 2. Data for each X molecule is stored in a python dict type containing all conformer data. The keys shown in figure 2: coordinates, energies, and species give access to containers of the type shown, and containing data described by the key. Species is a python list of strings containing the atomic symbol of each atom and its order corresponds correctly to dimension 1 of the coordinates numpy array. Appending "HE" to the end of the coordinates and energies keys will yield high energy structures as described in the technical validation section.

## Technical validation

Since normal mode sampling is used to generate the non-equilibrium structures, high-energy conformers exist in the data set. These high energy conformations occur where the harmonic approximation of normal modes fail in anharmonic regions of a potential, and are caused by atomic clashes or other highly unfavorable molecular conformations. The distribution shown in Figure 3b visualizes the energies in the dataset, which contains structures with energies as high as 15 Ha. For this reason, energies greater than 275kcal/mol higher than the lowest energy conformer were not included into the training set of the ANI-1 potential. This removed 2,630,435 (10.7% of the original total) structures yielding 22,057,374 structures. Regions this high in energy are generally not considered in bio-chemical research. However, this data might be useful for some purposes. Therefore, we include both the high-energy and the low energy datasets as described in the data description section. Figure 3c shows the new distribution of energies which are never larger than 0 Ha in total energy minus the sum of atomic contributions to the total energy.

During the structural optimization phase, we do not distinguish between optimized structures that might land at a saddle point in the potential surface and those that land at some structural minima. Given the goal of sampling conformational space, the fact that some structures might land at off equilibrium geometries (saddle points) could in fact help in using this data to fit potential surfaces, as it will help to cover regions of conformational space not covered by equilibrium molecule normal mode sampling. However, if the optimization fails to converge to a stationary point, as 485 molecules did, then these structures were not included in the training set, as the validity of their configuration could not immediately be confirmed. However, given the vast number of structures in the data set, it is likely any interaction found in these 485 molecules can be found elsewhere in the data set.

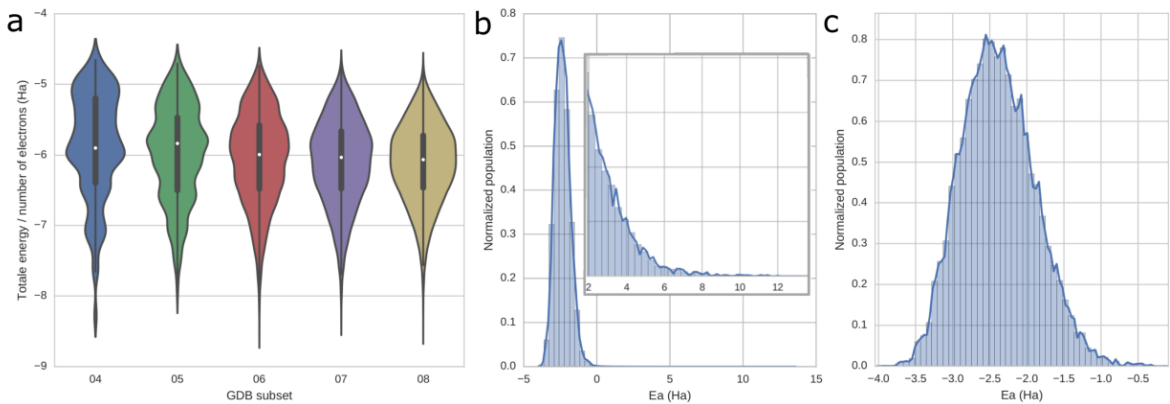

*Figure 3*: **Dataset energy distribution.** a) *The distribution of total energies divided by the number of electrons from normal mode sampling conducted on each sub set (04 through 08) of GDB-11. Each distribution is scaled to have equal area. b) Distribution of atomization energies from the completed data set with the inset showing a long tail reaching greater than 12 Ha. c) Distribution of atomization energies after truncating any energies over 275kcal/mol from each molecule's minimum energy.*

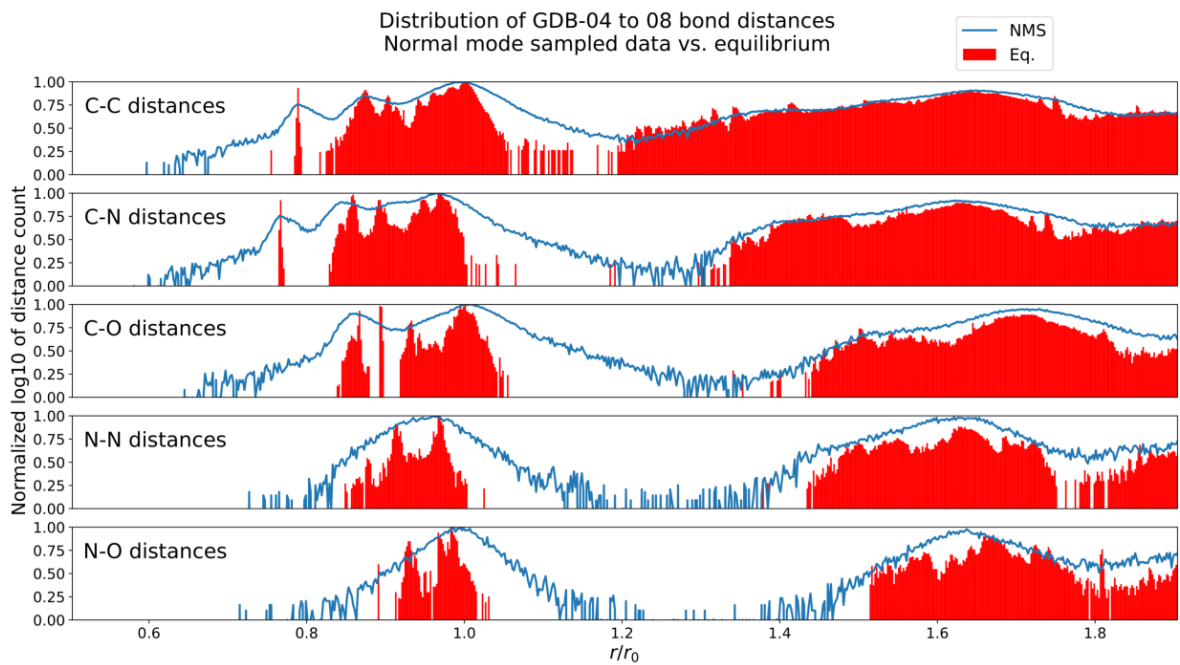

*Figure 4*: ***Distance distribution for dataset.*** *Distribution of atomic distances in the subset of the data set constructed from the molecules containing between 4 and 8 heavy atoms (GDB-04 to 08) of C, N, and O. The y-axis is the base 10 logarithm of the count of distances in each bin, normalized over the full domain so that the two sets can be compared. The x-axis represents the atomic distance (r) divided by the single bond equilibrium distance ($r_0$) for the smallest possible molecule containing a single bond of the type shown, as calculated using the ωB97x density functional with the 6-31g(d) basis set. The red histogram shows the full distribution of distances for a data set containing only equilibrium distances. The blue line shows the distribution of our non-equilibrium data set, with distances randomly sub sampled at a rate of 1%. As the figure shows, even 1% of the non-equilibrium data set covers vast areas of atomic distance space where the equilibrium data set fails to sample.*

A similar process of not including information for unconverged calculations is used in the generation of the total energies. For certain highly elongated bonds the molecular orbital optimization process, the self-consistent field procedure used in obtaining the total energy of the conformation, can fail to converge to a solution when two orbitals are too close in energy. For this reason, if a structure's single point energy calculation failed to converge, then this data is not included in the data set.

The primary concept of including non-equilibrium data is to sample regions of chemical space that would be sparsely covered in equilibrium only data sets. Figure 3a provides validation of energy sampling by showing the distribution of total energies divided by the total number of electrons for each molecule in the GDB subsets from 4 to 8-heavy atoms. Figures 3b and 3c show the distribution of total energies minus the sum of all individual atomic energies (tabulated in SI Table S1) for the full and "low energy" (less than 275kcal/mol from the minimum energy) data sets, respectively.

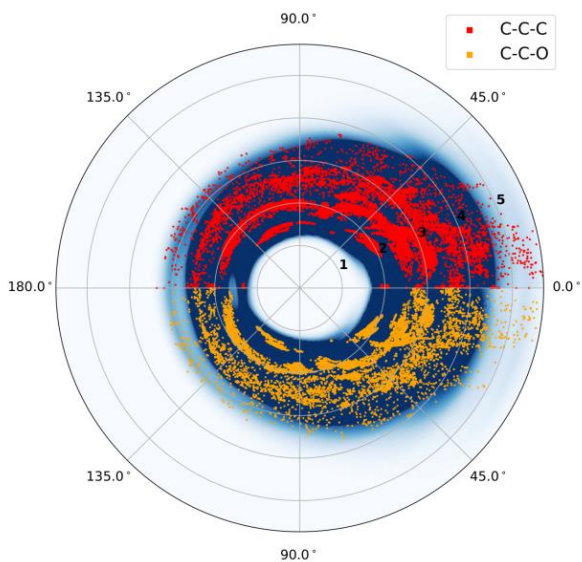

*Figure 5: Angular distribution. Figure shows distributions involving the angles in the data sets, and tells a similar story in terms of coverage in conformational space for three body interactions. The blue background density plot shows that the ANI-1 data set better covers angle space than the equilibrium data sets (red and orange). The remaining figures for the angular distributions are included in the SI.*

Further validation of the of non-equilibrium sampling is to show the data set covers a large domain of the chemical degrees of freedom in conformational space. Figure 4 contains five panels representing the distribution of atomic distances in the resulting non-equilibrium data set (blue line) compared with a data set of equilibrium only conformations (red) of the same molecule. As expected, the normal mode sampling method used to generate non-equilibrium conformations visits areas of conformational space not covered by equilibrium only data. A similar plot, SI Figure S1 shows distance distributions for the remaining atomic pairs. Figure 5 shows distributions involving the angles in the data sets, and tell a similar story in terms of coverage in conformational space for three body interactions. The blue background density plot shows that the ANI-1 data set covers far more angular space than the equilibrium data sets (red and orange). The remaining plots are included in the SI Figures S2 to S4.

## Usage Notes

To ensure that all readers have easy access to the ANI-1 data set, we have developed a python library with an easy to use interface for extracting the data. Examples uses of this library are included in the "readers" folder.

## Acknowledgments

J.S.S. acknowledges the University of Florida for funding through the Graduate School Fellowship (GSF). A.E.R. thanks NIH award GM110077. O.I. acknowledges support from DOD-ONR (N00014-16-1-2311) and Eshelman Institute for Innovation award. Part of this research was performed while O.I. was visiting the Institute for Pure and Applied Mathematics (IPAM), which is supported by the National Science Foundation (NSF). The authors acknowledge Extreme Science and Engineering Discovery Environment (XSEDE) award DMR110088, which is supported by National Science Foundation grant number ACI-1053575. We gratefully acknowledge the support of the U.S. Department of Energy through the LANL/LDRD Program for this work. We also acknowledge Nicholas Lubbers and Roman Zubatyuk for stimulating discussions and technical help with data organization.

## Data Citation

1. Smith, J. S., Isayev, O., Roitberg A. E. *Figshare*. https://figshare.com/s/2bdb6c25a547643d3db8 (2017)

# Supplementary information for: "ANI-1, A data set of 20 million calculated off-equilibrium conformations for organic molecules"

Justin S. Smith[1], Olexandr Isayev[2,*], Adrian E. Roitberg[1,*]

[1]*Department of Chemistry, University of Florida, Gainesville, FL 32611, USA*

[2]*UNC Eshelman School of Pharmacy, University of North Carolina at Chapel Hill, Chapel Hill, NC 27599, USA*

\* Corresponding authors; email: O.I. (olexandr@olexandrisayev.com) or A.E.R. (roitberg@ufl.edu)

| Atomic Species | Self-interaction energy |
|---|---|
| H | −0.500607632585 |
| C | −37.8302333826 |
| N | −54.5680045287 |
| O | −75.0362229210 |

Table S1: Self-interaction energies of atoms. Calculated with the ωB97x functional with the 6-31G(d) basis set. Each atom is treated with the proper spin state for the neutral species.

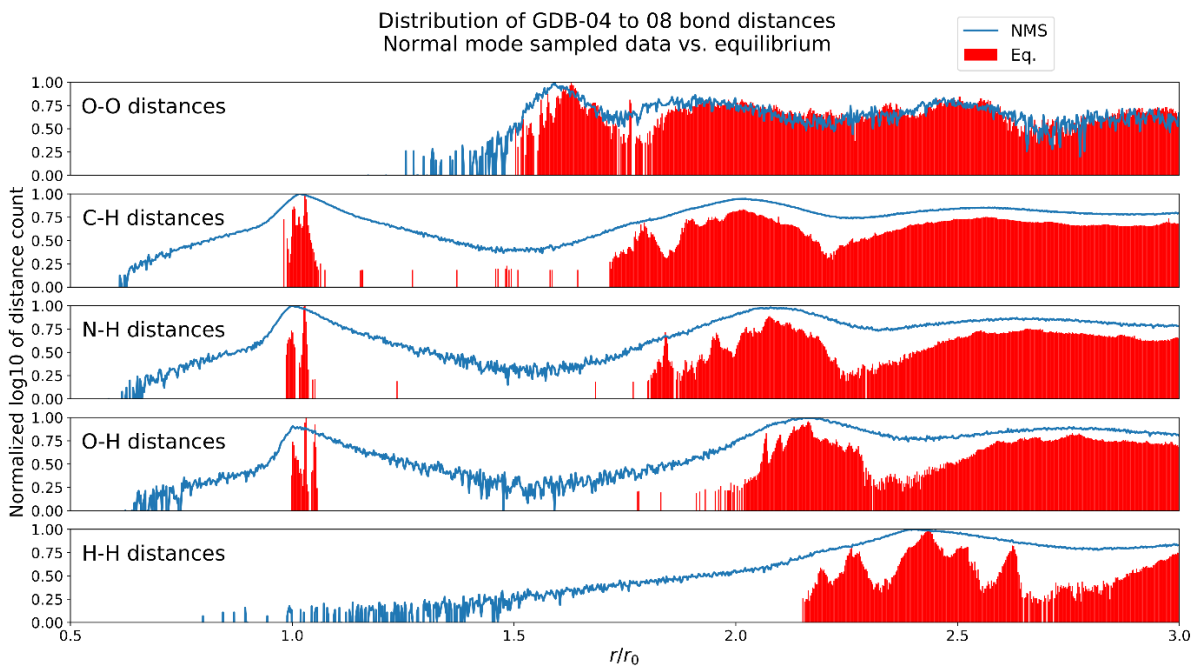

*Figure S1:* *Distribution of atomic distances in the subset of the data set constructed from the molecules containing between 4 and 8 heavy atoms (GDB-04 to 08) of C, N, and O. The y-axis is the base 10 logarithm of the count of distances in each bin, normalized over the full domain so that the two sets can be compared. The x-axis represents the atomic distance (r) divided by the single bond equilibrium distance ($r_0$) for the smallest possible molecule containing a single bond of the type shown, as calculated using the ωB97x density functional with the 6-31g(d) basis set. The red histogram shows the full distribution of distances for a data set containing only equilibrium distances. The blue line shows the distribution of our non-equilibrium data set, with distances randomly sub sampled at a rate of 1%. As the figure shows, even 1% of the the non-equilibrium data set covers vast areas of atomic distance space where the equilibrium data set fails to sample.*

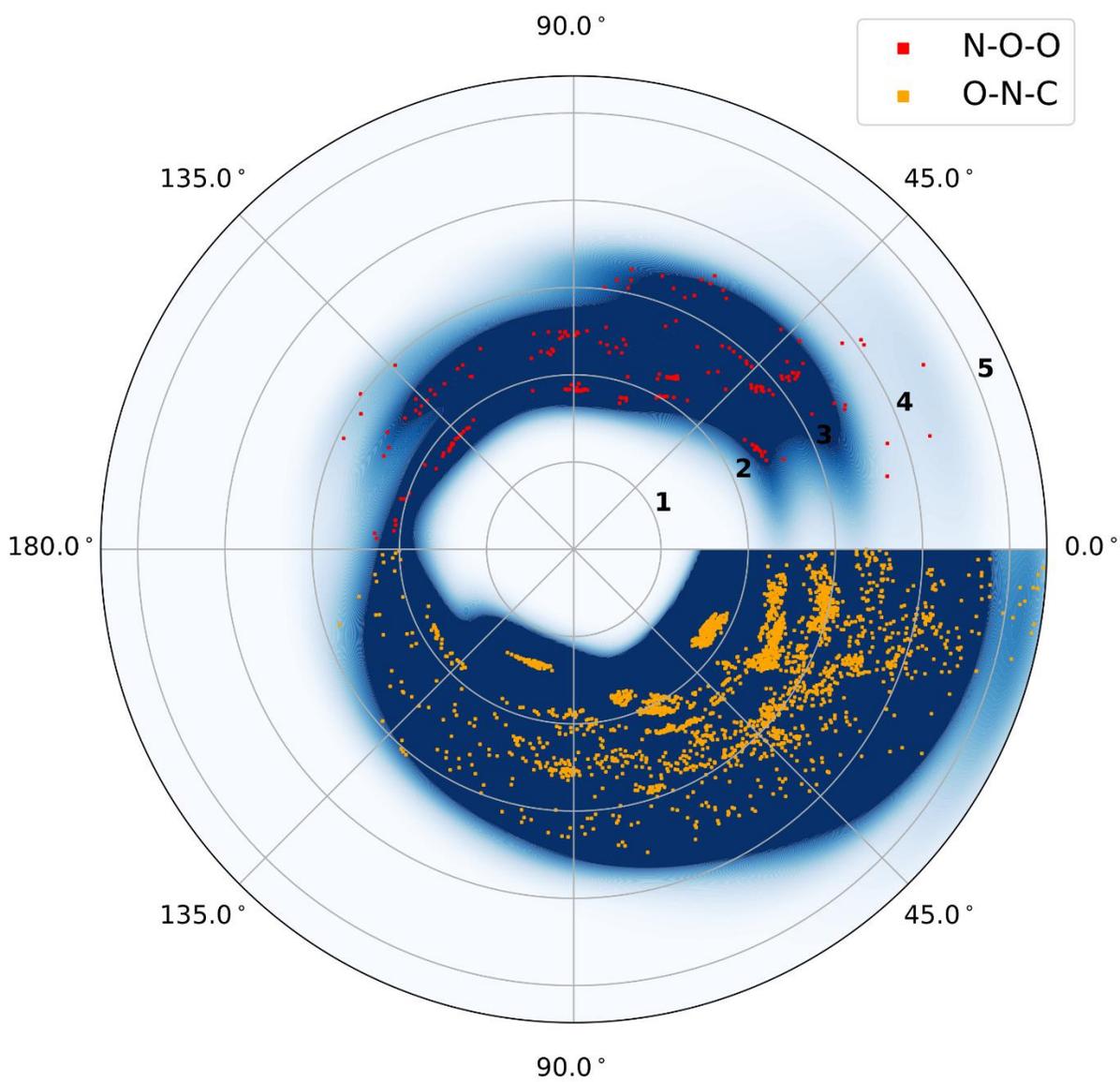

*Figure S2*: *Distribution of angles centered on carbon atoms and average distances between the other atoms and the carbon center. Red: N-O-O and orange: O-N-C atom triples for the equilibrium data set of 6-heavy atom molecules. The blue density plot in the background is from the 6-heavy atom non-equilibrium data set subsampled at 10% of all angles in the data set and saturated at 10% density.*

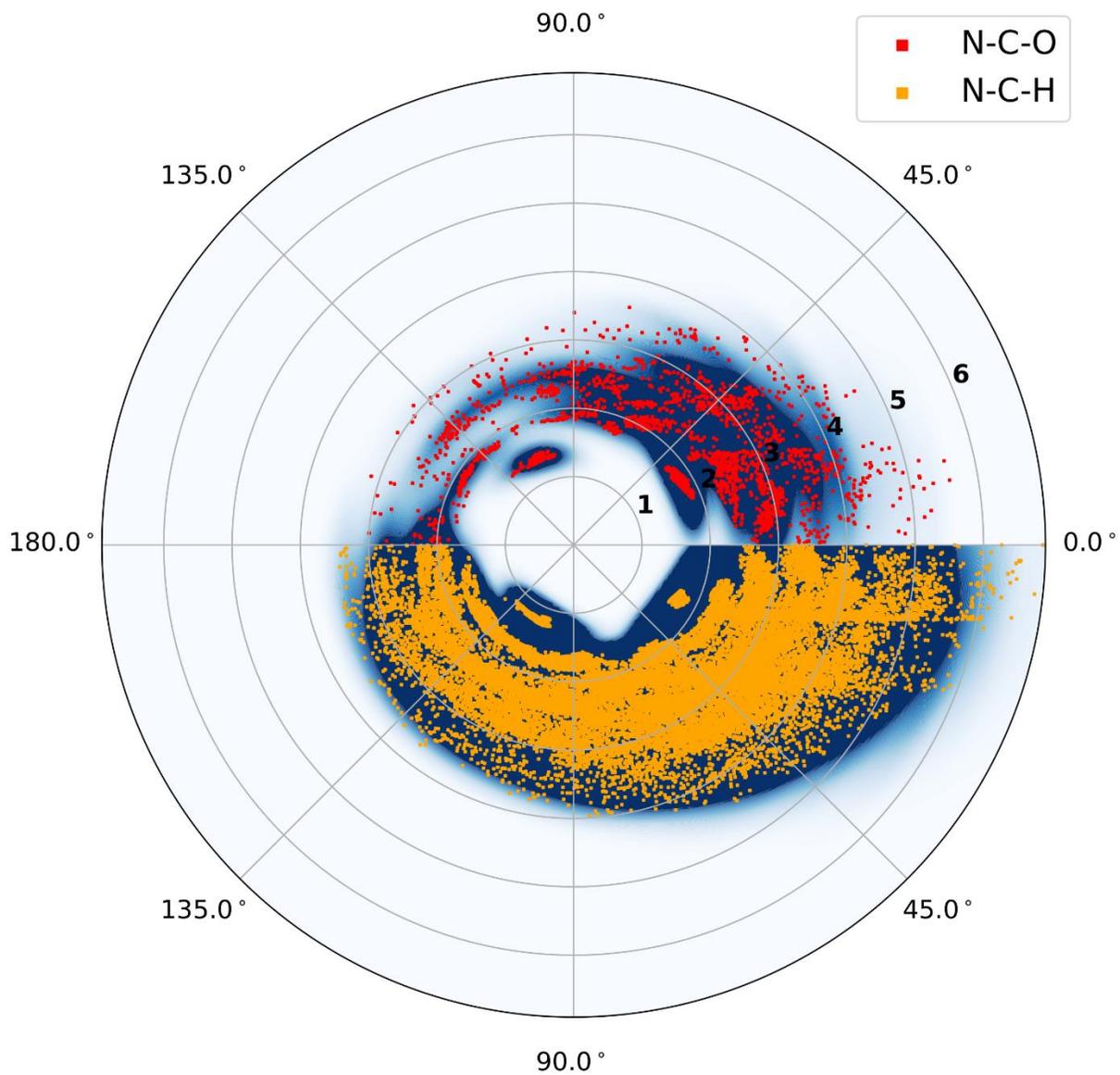

*Figure S3: Distribution of angles centered on carbon atoms and average distances between the other atoms and the carbon center. Red: N-C-O and orange: N-C-H atom triples for the equilibrium data set of 6-heavy atom molecules. The blue density plot in the background is from the 6-heavy atom non-equilibrium data set subsampled at 10% of all angles in the data set and saturated at 10% density.*

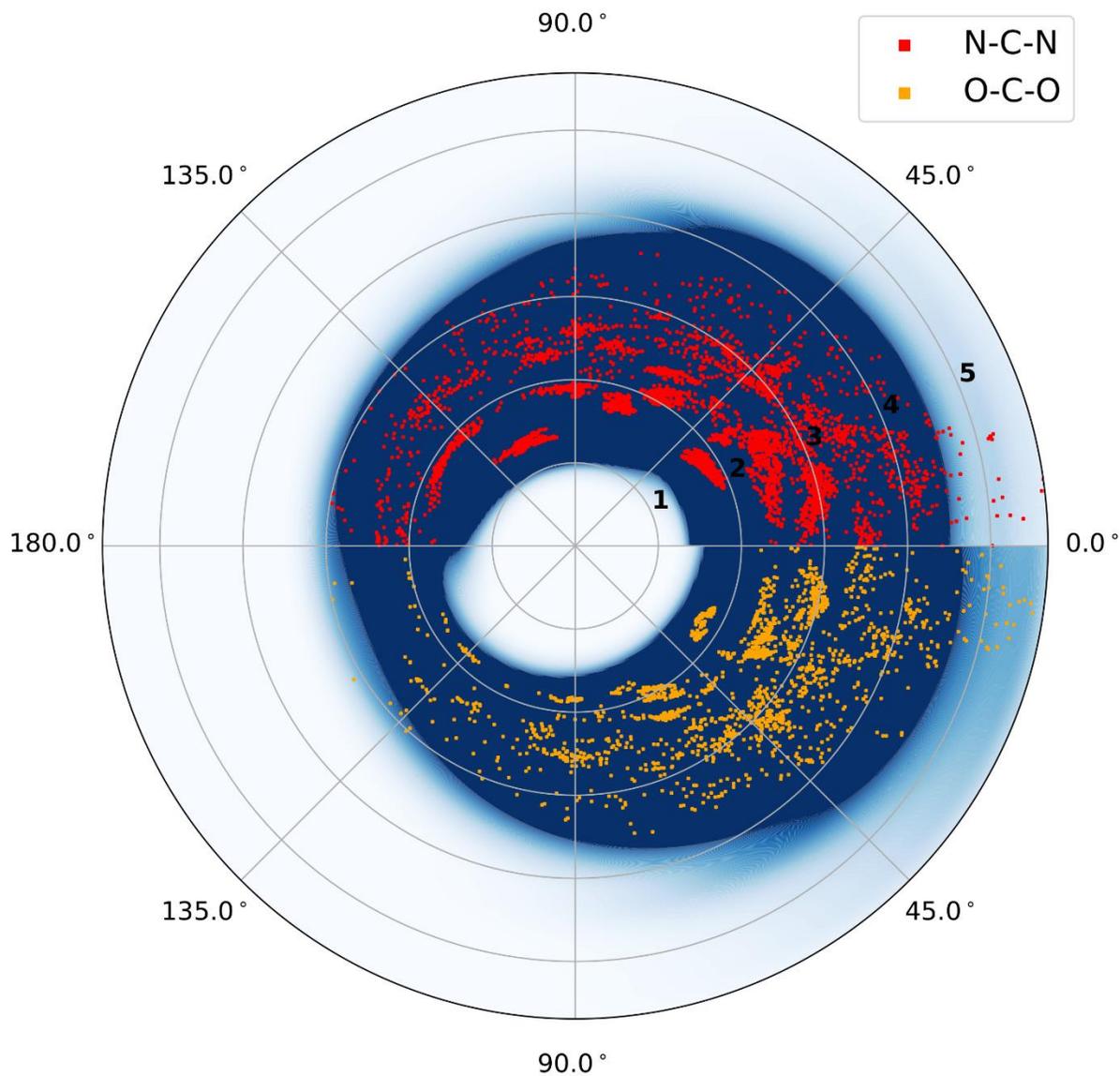

*Figure S4*: *Distribution of angles centered on carbon atoms and average distances between the other atoms and the carbon center. Red: N-C-N and orange: O-C-O atom triples for the equilibrium data set of 6-heavy atom molecules. The blue density plot in the background is from the 6-heavy atom non-equilibrium data set subsampled at 10% of all angles in the data set and saturated at 10% density.*